\documentstyle[11pt]{article}
\begin{document}

\newcommand \be  {\begin{equation}}
\newcommand \bea {\begin{eqnarray} \nonumber }
\newcommand \ee  {\end{equation}}
\newcommand \eea {\end{eqnarray}}

 \def\(({\left(}
 \def\)){\right)}
\def\[[{\left[}
\def\]]{\right]}
\def\bi{\bibitem}
\def \la{\langle}
\def\ra{\rangle}
\def \a{\gamma}
\def\ov{\over}
\def\l{\left}
\def\r{\right}
\def\b{\beta}
\def\de{\delta}
\def\p{\partial}
\def\D{\Delta}
\def\eps{\epsilon}
\def\lam{\lambda}
\def\hl{{\hat \lambda}}

\def \v{ {\vec v}}
\def \vx{\vec x}

\title{\bf VELOCITY FLUCTUATIONS IN FORCED BURGERS TURBULENCE}

\vskip 3 true cm

\author{Jean-Philippe Bouchaud$^1$ and Marc M\'ezard$^2$}

\date{\it
$^1$ Service de Physique de l'\'Etat Condens\'e,
 Centre d'\'etudes de Saclay, \\ Orme des Merisiers,
91191 Gif-sur-Yvette Cedex, France \\
$^2$ Laboratoire de Physique Th\'eorique de l'Ecole Normale Sup\'erieure
 \footnote {Unit\'e propre du CNRS,  associ\'ee
 \`a\ l'Ecole
 Normale Sup\'erieure et \`a\ l'Universit\'e de Paris Sud} , \\
24 rue
 Lhomond, 75231 Paris Cedex 05, France }


\maketitle

\begin{abstract}
We propose a simple method to compute the velocity difference
 statistics in forced
Burgers turbulence in any dimension. Within a reasonnable
assumption concerning the
nucleation and coalescence of shocks, we find in particular that the `left'
tail of the distribution decays as an inverse square power, which is compatible
with numerical data. Our results are compared to those of various
recent approaches: instantons, operator product expansion, replicas.

 \end{abstract}

\vskip 0.5cm

LPTENS preprint 96/37.

\vskip 0.5cm

\noindent Electronic addresses :
bouchaud@amoco.saclay.cea.fr, mezard@physique.ens.fr

\newpage
\section{Introduction}
Burgers equation, which describes the potential flow of a fluid
without pressure, provides a wonderful laboratory for testing new ideas and
techniques in view of the study of fully developped turbulence in the
Navier Stokes equation. These are two cases of non linear stochastic
equations which share the same structure of the non linearity. The
important difference comes from the nature of the large scale structures.
In the case of Burgers equation these are shock waves and the corresponding
physical picture of the flow is rather simple. This simplicity has already
allowed
for a very detailed study of the decaying turbulence \cite{Burg1d,esi}.

The
forced case, in which the fluid is stirred randomly and steadily on large
length scales, is more complicated. However it has been attacked recently
by various methods like the operator product expansion \cite{pol}, direct
probabilistic methods \cite{Kraich,Sinai}, instanton calculus\cite{gumi}
and replica method \cite{BMP}. The latter method allowed to get a detailed
solution in
infinite dimension, and the finite dimensional solution
seems to be within reach. This would be an important milestone for several
reasons.
It gives an example  of a flow with strong intermittency, created by large
scale structures.
It provides a benchmark to test new -or older- ideas on fully developped
turbulence.
Furthermore this problem is also related to interesting problems in condensed
matter
physics, like the elastic lines in random media (e.g. vortices in
superconductors),
and growth problems \cite{KPZ}. In this respect it is interesting to notice
that Burgers equation (with time playing the role of the running length scale)
also appears naturally in the renormalization group
study of manifolds in random media \cite{babome}. Therefore, the phenomenology
of
Burgers equation might be directly relevant to experimental studies of pinned
Bloch walls \cite{blochwalls} (or other elastic manifolds), besides its more
direct potential
applications to turbulence in one dimensional fluid flows \cite{kelrou} or to
pattern
formation in astrophysics \cite{zeldo}.

In this paper we shall focus on a simple aspect of forced Burgers turbulence:
the tails of the velocity gradient distribution in the regions where there are
no shocks.
We evaluate the `right' tail through a rather simple computation, and compare
it to the more sophisticated approaches developped recently
\cite{BMP,pol,gumi}. We then give a conjecture on the `left' tail which is
based on a plausible argument, requiring the system to reach a stationnary
state. We shall first discuss the one dimensional case,
then turn to higher dimensions, and compare our results with the previously
available ones.

\section{The slope dynamics in one dimension}

In one dimension we consider a velocity
field $v(x,t)$ governed by Burgers equation:
\be
{\partial v \over \partial t} + v  {\partial v \over  \partial x}
= \nu {\partial^2 v \over  \partial x^2} + f(x,t)
\ee
where $f(x,t)$ is a random force, which is supposed to have Gaussian
distribution,
with zero mean and a second moment:
\be
\la f(x,t) f(x',t') \ra= \eps \delta(t-t') R\(( {(x-x')^2 \over \D^2} \))
\ee
where $R$ is any smooth function decaying to zero fast enough at large
arguments (e.g. an exponential as in \cite{BMP}),
 $\D$ is the length scale of the stirring force and $\eps$ is
the injected energy density. To keep consistency with the notations of
our previous work \cite{BMP}, we fix the normalisations by requiring that
at short distance
$R(y)=1-\frac{3}{2} y+ O(y^2)$.

Our approach is based on the following observation: even in the forced case,
the
flow is organised in some smooth regions separated by shocks. Inside the smooth
regions, the flow locally depends linearly on the position:
\be
v(x,t) \simeq \lam(t) (x-x_0(t)) \ .
\ee
A linear expansion of the original equation allows to derive the evolution of
$x_0(t)$
(which is irrelevant for our present discussion) and of the local slope
$\lam(t)$. This
slope follows a Langevin equation:
\be
{d \lam \over dt} = -\lam^2(t) +\eta(t)
\label{langN1}
\ee
with a noise term $\eta$, due to the random stirring force, which has a
Gaussian
distribution, with:
\be
\la \eta(t) \eta(t') \ra = \frac{3 \eps}{ \D^2} \delta(t-t') \ .
\ee
This Langevin equation describes the relaxation of a particle at temperature
$T={3\eps \over 2\D^2} $ in a potential $V(\lam)=\lam^3/3$. The zero
temperature
case corresponds to the decaying Burgers turbulence, where the slope decays in
time as
$\lam(t)=\lam_0/(1+\lam_0 t)$, leading to an asymptotic behaviour in which the
slope is
$1/t$ independently of the initial conditions \cite{Burg1d}. The forced case $T
\ne 0$
leads to runaway solutions: starting from a positive slope $\lam_0$, it will be
eventually driven by the forcing to an infinitely negative slope, $\lam \to
-\infty$.
This effect is nothing but the building up of a new shock, which tends to
develop as
soon as the fluctuations due to the forcing drive the system to a negative
slope.
However the tail
 of the
slope distribution at positive $\lam$ is unaffected by this effect
(see below, Eq.(\ref{Pfull})): it depends
on rare noise configurations which, for a Gaussian forcing, lead to a Boltzmann
form:
\be
P(\lam) \sim C e^{-V(\lam)/T} = C \exp\((-{2 \D^2 \lam^3 \over 9 \eps}\))\ , \
\ \ \
(\lam \to + \infty)
  \ ,
\label{rightt}
\ee
where $C$ is a constant. A similar result was obtained in the context of
pinned Charge Density Waves in \cite{Feigelman}.

Inside the linear regions, the velocity difference between two points at
distance $r$,
$u=v(x+r)-v(x)$,
equals the slope $\lambda$ times the distance $r$, so that the slope
fluctuations
induce the following tail for the pdf of the velocity difference:
\be
P(u) \simeq \frac{C}{r} \exp\((-{2 u^3 \D^2 \over 9 \eps r^3}\)) \ ,\ \ \ \
( u \to + \infty)
\label{uright}
\ee
which is precisely the result derived in \cite{pol,gumi}. We shall discuss the
relationship
of this elementary derivation with these other more elaborate ones later on.

{}From the evolution of the slope (\ref{langN1}) one can also discuss, in a
more speculative way,
the other tail of the pdf of the velocity differences (at large negative $u$).
Since the regime we want to study is stationnary, the average number of shocks
(or of linear regions) must be conserved. Hence, the slopes which `disappear'
at $\lambda = -\infty$ creating a new shock and thus {\it two} new slopes, must
be compensated by the spontaneous disappearance of some other slopes through
shock coalescence. In other words, for the number of `cells' (locally linear
regions) to be conserved, one must introduce a source term $J_0 S(\lam)$ in the
corresponding Fokker-Planck equation, which describes these
`reinjection'/coalescence processes. These processes (and hence the correct
form of $S(\lam)$) are difficult to describe properly, but fortunately the
asymptotic results are rather insensitive to the detailed shape of $S(\lam)$.
We normalize the integral over $\lam$ of $S(\lam)$ to $1$, and write the Fokker
Planck equation for the Langevin process (\ref{langN1})
in the presence of the source term as:
\be
{\p P \over \p t}= T {\p^2 P \over \p \lam^2}+ {\p \over \p \lam} \(( P \lam^2
\))
+J_0 S(\lam) \ .
\label{PJ}
\ee
This equation now possesses a stationnary solution with a non vanishing current
$J_0$, which is nothing but the probability per unit time for a new shock to
appear. This stationnary solution is given by:
\be
P(\lam)= J_0\exp\((-{\lam^3 \over 3 T}\)) \int_{-\infty}^\lam  d\lam' F(\lam')
\exp\((+{\lam'^3 \over 3 T}\))\label{Pfull}
\ee
where $F(\lam)=\int_\lam^\infty  d\lam' S(\lam')$ and $J_0$ is fixed by
the normalisation of $P(\lam)$. If we make the reasonable
assumption that the reinjection term $S(\lam)$ is a rapidly decaying function,
the integral appearing in the above form (\ref{Pfull}) of the slope probability
converges for $\lambda \to +\infty$, leading back the above result, Eq.
(\ref{rightt}). For $\lambda \to -\infty$, $F(\lam) \to 1$, and the asymptotic
form of $P(\lam)$ is easily shown to be:
\be
P(\lam) \sim {J_0 \over \lam^2} \ , \ \ \
(\lam \to -\infty) \ .
\label{leftt}
\ee
This in turn leads to a velocity difference distribution which decays as:
\be
P(u) \sim {J_0 r\over u^2} \ , \ \ \
(u \to -\infty) \ .
\label{leftut}
\ee
which happens to be precisely the form that Chekhlov and Yakhot have proposed
using
very different arguments, and which is compatible with numerical data
\cite{CY}.

Let us emphasize that the Fokker-Planck equation without the source term
$S(\lam)$ does not have a (normalizable) equilibrium state. This normalizable
stationnary state can only appear in the presence of a non-zero current for
large
negative $\lam$. The absence of an equilibrium state is also at the heart of
the
operator product expansion approach of Polyakov \cite{pol}, on which we shall
come back below.

\section{The slope dynamics in higher dimension}

The generalization of this approach to the $N$-dimensional Burgers equation is
interesting. The velocity field verifies the equation:
\be
{\partial \v \over \partial t} + (\v \cdot \vec \nabla) \ \v
= \nu \nabla^2 \v + \vec f(\vec x,t) \
\ee
where $\v$ is a gradient flow and the forcing term $\vec f$ is  a random
gradient field,
with a Gaussian distribution of mean zero and second moment given by:
\be
\overline{f^j(\vec x,t) f^k(\vec x',t')} =
\epsilon \delta(t-t') \left[\delta^{j k} - {(\vec x - \vec x')^j
(\vec x - \vec x')^k \over N\Delta^2} \right] \
 G\left[{(\vx-\vx')^2 \over 2 N \Delta^2}\right] \ ,
\ee
where the $N$ dependance is chosen such as to insure the existence of
 a well defined large $N$ limit
\cite{BMP}. As before, the correlation $G$ decreases fast enough at large
arguments and behaves
as $G(y) \simeq 1-\frac{y}{2}+O(y^2)$ at small arguments.

Between the shocks (which now have a dimension $N-1$), the flow is locally
radial, with
a slope matrix $M$: \be
v_j(\vx,t)=\sum_k M_{jk}(t) (x_k-x_k^0(t)) \ .
\ee
The slope matrix evolves in time as:
\be
{d M_{jk} \over dt} = -\((M^2\))_{jk}(t) + y_{jk}(t) \ ,
\ee
where $y_{jk}$ is a Gaussian random noise with a zero mean and a variance given
by:
\be
\la y_{ij}(t) y_{kl}(t') \ra = {\eps \over N \D^2} (\delta_{ij} \delta_{kl}
+\delta_{ik} \delta_{jl}+ \delta_{il} \delta_{jk}) \delta(t-t')\ .
\ee
We shall need to study the statistics of the eigenvalues $\lam_i$ of the
slope matrix $M$. From second order perturbation theory, one finds that
these eigenvalues verify a Langevin type equation:
\be
{\p \lam_i \over \p t} = -\lam_i^2 + {\eps \over  N \D^2} \sum_{j(\ne i)}
{1 \over \lam_i-\lam_j} + \eta_i \ ,
\label{langN}
\ee
where the level repulsion term has appeared \cite{Dyson}, and the noise is
Gaussian
with a correlation:
\be
\la \eta_i(t) \eta_j(t') \ra = {\eps \over  N \D^2} \((1+2 \delta_{ij}\))
\delta(t-t') \ .
\ee
Compared to the one dimensional case, the situation is more complicated
because of two effects: level repulsion on one hand, and correlation of the
noises
$\eta_i$ acting on each eigenvalue on the other hand. This noise correlation
prevents the existence of an equilibrium (currentless) Boltzmann distribution,
which one could use as above to obtain the right tail of the slope
distribution. We shall
instead use an instanton computation, which is well suited to deal with such
correlations (see e.g. \cite{Bray}).

We want to compute the pdf $P(u)$ of the longitudinal velocity difference,
\be
u= \sum_i \((v_i(\vx+\vec r)-v_i(\vx)\)) {r_i \over |\vec r |} \ ,
\ee
which is given between the shocks by:
\be
u=r \sum_{ij} \hat r_i M_{ij} \hat r_j \ ,
\ee
where $r=|\vec r|$ and $\hat r$ is a unit vector in the direction of $\vec r$.
In terms of the eigenvalues of $M$, it takes the form:
\be
u=r  \sum_{i=1}^N \lam_i a_i^2
\ee
where $\vec a$ is a random unit vector (obtained from $\hat r$ through
a random rotation).

Let us introduce the Laplace transform of the pdf of the longitudinal velocity
difference:
\be
G(\mu,r)=\int du P(u) e^{\mu u} =\int_{-i \infty}^{i \infty} {dz
 \over 2 \pi i}
\int da_1...da_N \la \exp\((\mu r\sum_j a_j^2 \lam_j + z (\sum_j a_j^2-1) \))
\ra
\ee
In the spirit of recent works in turbulence \cite{giles,inst,gumi} and in other
fields
\cite{Bray,mader}, we shall use a path integral representation for the
probability
distribution of the eigenvalues, and evaluate
the large $\mu$ tail of
$G(\mu,r)$ by finding the leading
instanton trajectory. The  path integral is written through standard
manipulations
\cite{dedom}. The dynamics is started at a time $t_0 \to -\infty$, and the
eigenvalues
are measured at $t=0$. We introduce
conjugate fields $\hl_i(t)$ to implement the Langevin equation, and
get after average over the thermal noise:
\be
G(\mu,r)= \int_{-i \infty}^{i \infty} {dz \over 2 \pi i}
\int \prod_j da_j \int \prod_j  d[\lam_j] d[\hl_j] \exp\(( -S\))
\ee
with an action:
\bea
-S&= & \int_{-\infty}^0 dt \sum_j
\hl_j(t)\(( {d \lam_j \over dt} +\lam_j^2(t) -{\eps \over  N \Delta^2}
\sum_{k(\ne
j)} {1 \over \lam_j(t)-\lam_k(t)} \)) \\
& +&
\int_{-\infty}^0 dt {\eps \over 2 N \Delta^2}
\(( 2 \sum_j \hl_j(t)^2 + \sum_{j,k} \hl_j(t) \hl_k(t) \)) +
\mu r\sum_j a_j^2 \lam_j(t=0) + z \((\sum_j a_j^2-1\))
\label{action}
\eea
The stationnarity conditions of this action are easily written. The effect of
the
parameter $\mu$ on the trajectories $\lam_j(t),\hl_j(t)$ is just to impose the
boundary condition $\hl_j(t=0^-)=-\mu r a_j^2$. The action $S$ has the
following
scaling property: under the change $\mu \to \gamma^2 \mu$, $t \to \gamma t$,
$\lam_i \to \gamma \lam_i$, $\hl_i \to \gamma^2 \hl_i$ and $z \to \gamma^3 z$,
all the terms scale as $\gamma^3$, except the level repulsion term which is
left invariant. Therefore, in the large $\mu$ limit, it is legitimate to
neglect this level repulsion effect, and the following
instanton solves these simplified equations:
  \bea
a_2=...=a_N=\hl_2=...=\hl_N=0 \\ \lam_1(t)={\phi \over 1 -\phi t} \ \ \
\hl_1(t) = -{\mu r
\over (1-\phi t)^2}
\label{instN}
\eea
where the parameter $\phi$ is: $\phi=\sqrt{3 \eps \mu r/(2N
\D^2)}$. The
 action of this instanton is:
\be
-\ln(G(\mu,r)) \simeq S=- \sqrt{2 \eps \over 3 N \D^2} (\mu r)^{3/2}
\label{Gres}
\ee
Assuming that this instanton gives the leading contribution to the
Laplace transform $G$ at large
$\mu$, one deduces from (\ref{Gres})
the following tail of the
pdf of velocity differences at large $u$:
\be
P(u) \sim C \exp \((-{2 N \D^2 \over 9 \eps} {u^3 \over r^3} \))\qquad({u \to +
\infty}) \ .
\label{righttn}
\ee
In order to check the dominance of the instanton solution (\ref{instN}), we
have performed two checks: an evaluation of the contribution of other
instantons, and a numerical simulation of the Langevin equation.
Without level repulsion, other solutions can be
found, with $(N-p)$ $a_j$ and $\hl_j$ identically zero, and the $p$ remaining
$\lam$
and $\hl$ all equal. The action of these instantons is a factor
$f_p=\sqrt{\frac{p+2}{3p}}$ smaller than (\ref{Gres}). Since $f_p < 1$ for all
$p >
1$, these other solutions can all be neglected. Taking
into account the level repulsion can only increase further the action of these
other solutions, so they should not contribute.

A direct numerical test of the result (\ref{righttn}) is easily done. We have
simulated
the Langevin evolution equation for the eigenvalues (\ref{langN}) in the case
$N=2$.
We adopted a simple Ito discretization of time and computed the histogram of
$u/r =\lam_1 \cos^2(\psi)+ \lam_2 \sin^2(\psi) $ where $\psi$ is a random angle
with uniform
distribution in $[0,2\pi]$. The time step has been kept adaptive in order to
be able to use a small time step whenever the two eigenvalues are close to each
other.
Our simulation has been done at $\eps/\Delta^2=10/9$ and the statistics
is over $10^6$ points.
The result of the histogram is given in Fig.1, where the straight line gives
the
slope predicted from our instanton computation (\ref{righttn}). The agreement
confirms that the instanton (\ref{instN}) indeed gives the leading
contribution.
We have also simulated the Langevin process without the level repulsion term,
and checked that we find the same tail for $\ln P(u)$. Let us just mention
that the finite $u$
effects turn out to be larger in this latter case, and the analytic prediction
for the tail is found only in the regime where $\ln P(u) < -10$. This finite
$u$ corrections should also
be important for
the analysis of velocity difference tails in numerical simulations of Burgers
equation.

Altogether, we are confident that the right tail of the velocity difference is
given by (\ref{righttn}) also in the multidimensional case.
It is important to notice that this tail gets thinner and thinner as $N$ grows,
and
finally disappears in the limit $N \to \infty$, which was actually the
situation
that we considered in our previous paper \cite{BMP} using the replica method.
Indeed, the calculation that we
performed there shows that, for $N=\infty$,
$P(u)$ is strictly zero when $u > u_\D \frac{r}{\D}$
(where $u_\D=(\eps \D)^{1/3}$ is the velocity at scale $\D$). This is
compatible with Eq.
(\ref{righttn}) for $N = \infty$. It however clearly emphasizes the limitations
of a large $N$ approach: although many features of the velocity field are
correctly predicted, such a method is unable to predict the tails of the
velocity distribution. Stated differently, the limits of large velocity
difference and large dimension do not commute.

\section{Comparison with the direct instanton approach}

The previous results (\ref{rightt},\ref{righttn}) on the
tail of the longitudinal velocity difference at large positive $u$ are
identical to
those obtained in \cite{pol,gumi} in one dimension and in \cite{gumi} in
dimensions
$N>1$. The one dimensional result is thus confirmed by a very simple and direct
derivation. In higher dimensions, we are unfortunately
not able to solve fully for the probability distribution of the $N$ eigenvalues
because
of the noise correlations. We have rather found a simple instanton solution,
which
allows us to generalize the one dimensional result to arbitrary dimension.
 A naive generalisation of Eq. (\ref{PJ}) to higher dimensions
also suggest that the left tail of the distribution of velocity difference
should
decay as $u^{-2}$ in any finite dimension.

Although our instanton computation looks much easier than the one developped by
Gurarie and Migdal \cite{gumi} (we are dealing with the probability
distribution
of $N$ eigenvalues $\lam_i$ instead of a $N$ component velocity field
$\v_i(x)$), the two
methods are actually very similar. In one dimension for instance,
the instanton for the velocity field found in \cite{gumi} corresponds precisely
to a velocity field growing linearly with distance, with a slope called
$\sigma$ in \cite{gumi}.
It is easy to deduce from the equations (19,20) of \cite{gumi} an evolution
equation
for $\sigma$ which is identical to the one which is  derived from
the stationnarity of (\ref{action}). The same is true in larger dimensions.

To conclude our comparison with the work of \cite{gumi}, we have found that the
positive $u$ tail of $P(u)$ derived from their nice instanton solution
is certainly correct at least in one dimension. Our
computation also indicates that this instanton cannot directly give the tail
at negative $u$. In fact its structure, with a linear velocity field,
shows that this instanton takes into account basically
the linear regions, and not the shocks. As we argued before, these
linear regions dominate the right tail of $P(u)$, but the left tail
is given by a completely different process. In order to understand this
left tail quantitatively, one needs to
 control both the formation of shocks, and the statistics of the sizes of the
jumps in
velocity at the shocks (or equivalently the statistics of the lengths of the
linear
regions). This has been achieved in the decaying problem \cite{Burg1d}, or in
the
forced case in large dimensions \cite{BMP}, and it is hard to think that this
information is contained in a simple linear instanton configuration.

\section{Comparison with the operator product expansion}

Let us now discuss the operator product expansion (OPE) approach of Polyakov
\cite{pol},
which is supposed to give the {\it full} pdf $P(u)$ in the scaling region.
Using
some conjecture on the structure of the fusion rules in the OPE,
one finds that, in one dimension, this $P(u)$ is exactly the totally
assymmetric L\'evy stable
distribution of index $\frac{3}{2}$, given by the inverse Laplace
transform of the function $G(\mu,r)$ found in (\ref{Gres}).
 The behaviour of the right tail of $P(u)$ is
again given by Eq. (\ref{rightt}), and is thus compatible with the other
approaches. However one should emphasize that this tail does not constitute a
test for
the fusion rules of the OPE.
In the simple case of the `two point' function $P(u)$, the OPE approach relies
on the following steps. Restricting here to the case $N=1$, one first
introduces
 a generating functional $Z(s_1,s_2,t)\equiv \langle \exp[s_1 v(x_1,t) + s_2
v(x_2,t)] \rangle$. In the stationnary regime, one should have
$\frac{\partial Z}{\partial t}=0$, which, using the tricks of ref.\cite{pol},
can also be written as:
\be
\frac{\partial^2 Z}{\partial s_1 \partial x_1} +
\frac{\partial^2 Z}{\partial s_2 \partial s_2} -
\frac{\partial Z}{s_1 \partial x_1} -
\frac{\partial Z}{s_2 \partial x_2} = \langle [s_1 f(x_1,t) + s_2 f(x_2,t)]
\exp[s_1 v(x_1,t) + s_2 v(x_2,t)] \rangle +{\cal A} \ ,
\ee
where $\cal A$ is a term coming from the viscosity contribution, which is
singular in
the $\nu \to 0$ limit (the ``anomaly").
Using the fact that the force $f$ is Gaussian, the first term in the right
hand side can be reexpressed
as:
\be
\frac{Z \eps}{2} [R(0) (s_1^2 + s_2^2) + 2R(x_1-x_2) s_1 s_2]
\ee
Now, changing variables to:
\be
x_1=x+\frac{r}{2} \quad x_2=x-\frac{r}{2} \quad \mu=\frac{s_1-s_2}{2}
\quad \mu'=\frac{s_1+s_2}{2}
\ee
and using translational invariance, one finds, for $\mu'=0$, the following
final equation for $G(\mu,r)=\langle \exp[\mu (v(\frac{r}{2})- v(-\frac{r}{2})]
\rangle$:
 \be \(( {\p \ov \p \mu} -{2 \ov \mu} \))
{\p \ov \p r} G = {3 \mu^2 \eps \ov 2 \D^2} r^2 G +{\cal A} \label{poldif}
\ee
where $r << \D$ is assumed.  The simplest fusion rule leads to a
contribution scaling as
\be {\cal A} \simeq \alpha G +{\beta \ov \mu} {\p G \ov \p r}
\label{anomaly} \ee
where $\alpha$ and $\beta$ are two constants to be determined. It turns out
that the
right hand tail of the solution $G(\mu,r)$ of (\ref{poldif}) has a leading
behaviour
$\ln G \simeq \sqrt{2\eps/(3\D^2)} (\mu r)^{3/2}$ which is independent both of
$\alpha$ and $\beta$ !
Therefore
the $\exp(-c u^3)$ tail comes out exactly as what we derived in (\ref{uright})
(with precisely the same $c$), independently of the fusion rule. This is
actually
expected since we argued that this tail comes from the linear regions where one
can
completely neglect the anomaly. Without anomaly, the above equation has no
solution which can be interpreted as the Laplace transform of a probability
distribution (cf. the discussion after Eq. (\ref{leftut})). The fusion rules
thus
become crucial when one looks at the negative $u$ tail. If one insists that the
solution $G(\mu,r)$ of (\ref{poldif}) is the Laplace transform of a
(normalizable)
probability law, with an anomaly given by (\ref{anomaly}) with $\alpha=0$ for
all $\mu$, one
finds that $P$ is a L\'evy distribution of index $\frac{3}{2}$, with a left
tail
decaying as $P(u) \sim |u|^{-5/2}$ \cite{pol}. This result
disagrees with our above calculation, which suggested a $|u|^{-2}$ tail from
the
existence of a finite outgoing current at large (negative) slopes, related
to the formation of new shocks. We do not have enough control on this
`reinjection process' of the slopes  to make any strong statement. However,
our argument is suggestive enough to try to look at other fusion rules which
indeed
lead to such a tail \cite{polprivate}.

\section{Conclusion}

We have discussed the tails of the velocity difference distribution in Burgers'
forced turbulence. We have shown how a direct method gives result which are in
agreement with other methods in the right tail of the distribution. This method
however suggests that the left tail is governed by the dynamics of shock
formation,
and requires some control over the shocks nucleation/coalescence processes. A
simple
hypothesis describing these processes in terms of an effective source term in
the
corresponding Fokker-Planck equation gives a $|u|^{-2}$ tail for the left side
of the
distribution, at variance with the simplest OPE conjecture, but apparently
in agreement with numerical simulations.

It would be very interesting to extend these methods to describe more
complicated
quantities, such as the statistics of the velocity potential $h$ defined as
$\vec v = -
\vec \nabla h$. In particular, the statistics of the `barrier heights'
separating
the different valleys of $h$ (corresponding to the cells of the Burgers flow)
would
be of direct interest to understand in detail the long time dynamics of
randomly pinned
objects{\footnote{Conversely, as originally noticed by Feigel'man
\cite{Feigelman},
the very large `slope' tail $\lambda \to \infty$ (Eq. (\ref{rightt})) of the
Burgers
flow is related to the high frequency response of these pinned objects}}. In
particular, it would be important to see to what extent the large $N$ result,
which
states that these barriers are exponentially distributed, has to be modified
for
finite $N$. Another interesting perspective is the generalization of these
approaches to the case of  a forcing correlated on long distances, where
Kolmogorov's scaling is observed between the shocks \cite{CYlr,hayot}.
\vskip 2cm
Acknowledgments. We want to thank A. Chekhlov, T. Dombre, V. Gurarie
 and V. Yakhot for interesting discussions.

{\bf Figure Captions}
\begin{itemize}

\item[Fig.~1] The logarithm of the probability distribution function of
velocity differences, $\ln P(u)$, versus  $(u/r)^3$, in dimension $N=2$,
 for $\eps/\Delta^2=10/9$.
The curve is obtained from the simulation of the
Langevin equation on the slope matrix eigenvalues, and the straight
line indicates the prediction from the
instanton computation (\ref{righttn}).
\end{itemize}
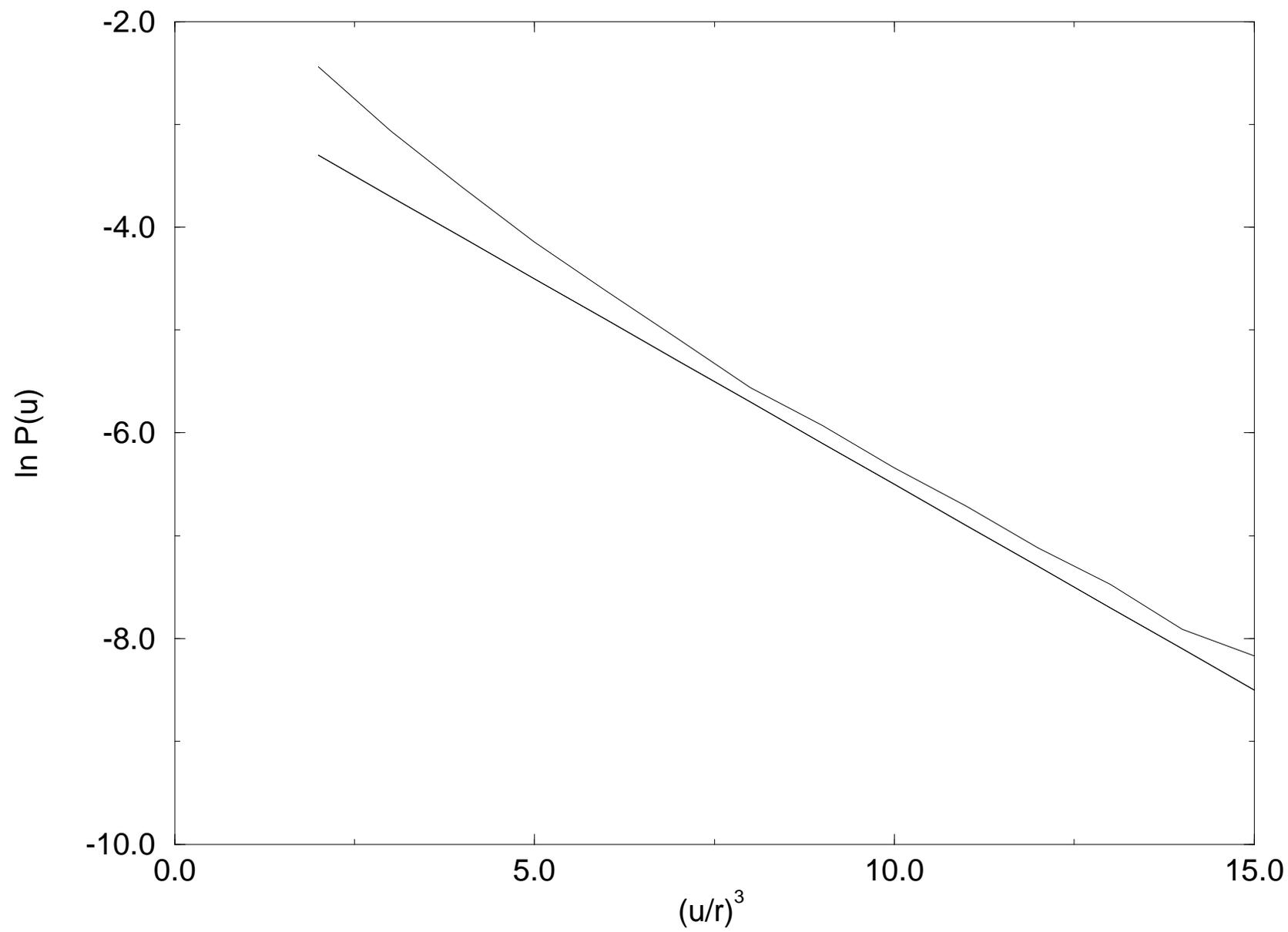
\end{document}